\documentstyle[prb,aps,psfig]{revtex}
\def\be{\begin{equation}}
\def\ee{\end{equation}}
\def\bc{\begin{center}}
\def\ec{\end{center}}
\begin{document}
\title{
Critical phenomena at perfect and non--perfect
surfaces}
\author{M. Pleimling and W. Selke}
\address{Institut f\"ur Theoretische Physik B, Technische Hochschule,
D--52056 Aachen, Germany}
\maketitle
 


\begin{abstract}
The effect of imperfections on surface critical properties is
studied for Ising models with nearest--neighbour ferromagnetic
couplings on simple cubic lattices. In particular, results of
Monte Carlo simulations for flat, perfect surfaces are compared
to those for flat surfaces with random, 'weak' or 'strong', 
interactions between neighbouring spins in the surface layer, and for
surfaces with steps of monoatomic height. Surface critical exponents
at the ordinary transition, in particular $\beta_1 = 0.80 \pm 0.01$,
are found to be robust against these perturbations.
\end{abstract}
 
\section{Introduction}
Critical phenomena at perfect, flat surfaces of three--dimensional crystals
have attracted much interest. In particular, surface critical
exponents have been estimated, both theoretically and experimentally.
The general agreement is quite satisfactory [1-3].\\
Nevertheless, there are still some open questions. For example,
the critical exponent $\beta_1$ of the surface magnetization
at the ordinary transition in the Ising universality class
seems to be known only to an accuracy of about ten percent, with
$\beta_1$ $\approx$ 0.80 $\pm$ 0.05, obtained
from large scale Monte Carlo simulations [4,5], renormalization
group calculations [6,7], and experiments on magnets and alloys
such as FeCo [3,8].\\
In addition, the role of surface imperfections, unavoidable in real
materials, on critical properties has been studied
for three--dimensional systems
in much less detail (for exceptions, see, e.g., mean--field theory [9] and
renormalization group arguments [10] on surface layers with
randomness), especially in simulations.\\
The aim of this article is twofold. Firstly, the values of critical
surface exponents for perfect surfaces at the ordinary transition
in Ising systems
have been reexamined by computing effective exponents allowing
to monitor easily the approach to the true asymptotic behaviour
and to refine previous estimates on the asymptotic exponents, e.g.,
on the value of $\beta_1$. 
Secondly, we have also performed Monte Carlo simulations for
two types of imperfections, corresponding to simple cases of
amorphous and corrugated surfaces, see Fig. 1. The amorphous surface is 
mimicked by choosing randomly 'weak'
or 'strong' nearest-neighbour ferromagnetic
couplings between spins in the flat surface layer. A corrugation is   
introduced in a simple manner by placing a terrace of
monoatomic height on
half of the surface; the magnetization at the step edge is
expected to reflect most strongly this form of 
perturbation as compared to the magnetization at the perfect 
surface.\\
The article is organized accordingly, concluded by a brief summary.
 
\section{Ising model with a perfect surface}
To simulate the critical behaviour of the perfect surface,
we study Ising models
on simple cubic lattices with free boundary conditions for the top
and bottom (surface) layers and periodic boundary conditions otherwise.
The spins are denoted by $S_{xyz}$ ($\pm$ 1), situated at lattice
sites $(x,y,z)$, where $z$ numbers the layers, $z= 1, 2,..., M$, with
$K \times L$ spins per layer ($x= 1, 2, ..., K$; $y= 1, 2, ..., L$).
Interactions are restricted to nearest--neighbours. Two different, 
ferromagnetic couplings may occur, depending on whether the neighbouring
spins are at the surface ($z =1$ or $M$), $J_s > 0$, or not, $J_b > 0$,
see Fig. 1a.\\
The phase diagram of the corresponding semi--infinite ($K, L,
M \longrightarrow \infty$ )
Ising model is well established [1,2]. If the ratio of the surface
coupling $J_s$ to the bulk coupling $J_b$, $r= J_s/J_b$, is 
sufficiently weak, the system undergoes at the bulk critical
temperature, $k_B T_c$/$J_b$ $\approx$ 4.5115 [11,12], an
'ordinary transition', with the bulk and surface ordering occurring at the
same temperature, $T_c$. Beyond a critical ratio, $r > r_c$ $\approx$
1.50 [5], the surface orders at a higher temperature, $T_s > T_c$,
followed by the 'extraordinary transition' of the bulk at $T_c$, see
Fig. 2. At the critical ratio $r_c$, one encounters the 'special transition
point', with critical properties of the surface transition deviating from
those at the ordinary or the distinct surface transition.\\
The aim of our simulations has been to reexamine the surface critical
properties at the ordinary transition, because of the rather wide
spread of supposedly accurate and reliable estimates of the critical
exponents, especially of the exponent of the surface
magnetization, $\beta_1$.\\ 
To study both surface and bulk properties, we computed various quantities, 
among others  the profiles of the
magnetization per layer, $m(z)$, and the susceptibility per
layer, $\chi(z)$, defined by\\ 
 
\be
m(z)= \frac{1}{K L} \left< \left| \sum\limits_{xy}  S_{xyz} \right| \right>
\ee

\noindent
and

\be
\chi(z)= \frac{K L}{k_B T} \left[ \left< \left( \frac{1}{K L} \sum\limits_{xy} S_{xyz}
\right)^2 \right> - \left( m(z) \right)^2 \right].
\ee

\vskip 0.5cm
\noindent
For $z= 1$(or $M$), one obtains the standard surface magnetization
$m_1 = m(z= 1)= m(z= M)$ and the susceptibility $\chi_{11}= \chi(z= 1)=
\chi(z= M)$. $\chi_{11}$ describes the response of the surface 
magnetization to a surface field [1,2]. The response of $m_1$ to a
bulk field is obtained from $\chi_1$ (which we also computed), given by [13]

\begin{equation}
\chi_1  =  \frac{K L M}{k_B T} \left[ \left< \frac{1}{2 K L} \left| \sum\limits_{surfaces}
S_{xyz} \right| \frac{1}{K L M} \left| \sum\limits_{sites} S_{xyz} 
\right| \right> 
- m_1 \, 
\frac{1}{K L M} \left< \left| \sum\limits_{sites}  S_{xyz} \right| \right>
 \right]
\end{equation}

\vskip 0.5cm
\noindent
Most of the Monte Carlo simulations were performed using the one--cluster
flip algorithm [14], augmented by runs using single--spin flips. System
sizes ranged from $K\times L\times M = 25\times 25\times 20$ to
$150\times 150\times 120$, thereby attempting to avoid finite--size
effects on approach to the phase transition; i.e. computing properties of
the semi--infinite system. To equilibrate the system and to extract
thermal averages, we usually generated a few $10^5$ clusters, and
determined error bars from typically 5 realizations.
Various ratios $r= J_s/J_b$ of the surface
to the bulk couplings were chosen, in between $r= 0$ and 2.0, studying 
predominantly the ordinary transition at $r= 1.0$.\\
Typical magnetization profiles are depicted in Fig. 3, at $r= 1.0$ and
$T < T_c$. The magnetization increases from its
surface value, $m_1$, to the bulk value $m_b$, at distances superceeding 
the bulk correlation length. To avoid finite--size effects,
$K = L$ and $M$ have to be sufficiently large. Firstly, $M$ may be chosen 
such that the profile $m(z)$
displays a pronounced plateau around the center of the system. Monitoring
then the $K$--dependence of $m_1$, one may determine the suitable size
of the Monte Carlo system. In Fig. 3, we included also the value
of the bulk magnetization, as had been obtained in extensive
and highly accurate previous simulations [12], thereby giving another check
on the quality of the
present data.\\
In order to estimate the critical exponents of the magnetization, we
define an 'effective exponent' $\beta_{eff}(z,t)$ by

\be
\beta_{eff}(z,t)= d \ln (m(z))/d \ln (t)
\ee

\noindent
where $t= |T-T_c|/T_c$ is the reduced temperature. Certainly, on approach to
$T_c$, $t \longrightarrow 0$, $\beta_{eff}$ becomes the asymptotic
exponent $\beta(z)$, with $\beta(z=1) =\beta_1$ being the critical 
exponent of the surface magnetization.\\

\noindent
Results for $\beta_{eff}(z,t)$ are shown in Fig. 4, at fixed temperatures,
$t$, and in Fig. 5, at fixed distances from the surface, $z$. Because the
simulations are performed at discrete temperatures, $t_i$, $\beta_{eff}$ 
is replaced by 

\be
\beta_{eff}(z,t)= \ln (m(z,t_i)/m(z,t_{i+1}))/\ln (t_i/t_{i+1})
\ee

\noindent
with $t =(t_i + t_{i+1})/2$. Very accurate Monte Carlo data are required
to get reliable estimates. Error bars may be assigned to the effective
exponents in various ways. We adopted the
form $\delta \beta_{eff} =(|\delta
m(z,t_i)/m(z,t_i)| + |\delta m(z,t_{i+1})/m(z,t_{i+1})|)$, where
$\delta m(z, t_i)$ follows
from the variance of the magnetization data due to the different
realizations, i.e. Monte Carlo runs with different random
numbers. These error bars are rather conservative as compared,
for instance, to those resulting from standard error
propagation assumptions.\\
From Fig. 5, one sees that the effective exponent of the surface
magnetization $\beta_{eff}(z=1)$ increases almost linearly with $t$ over
a wide range of temperatures. Note that we omitted, for clarity, data
which are obviously affected by finite-size effects.
A linear extrapolation to the critical
point yields $\beta_1 = 0.80 \pm 0.01$. This value is expected to correct
and refine the  previous estimates [3-8], ranging from 0.75 [5] to
0.845 [7]. It is nicely consistent with the value estimated from
a log-log plot for simulational data in between $0.01 < t < 0.1$,
with $\beta_1 =0.78 \pm 0.02$ [4] (being, presumably, slightly 
too low, because
the rise of $\beta_{eff}$ with smaller $t$ had been neglected).\\
In the non--asymptotic region, the surface magnetization may be
cast in the form $m_1 \approx m_0 t^{\beta_1} (1 + a t^x)$. From the data
depicted in Fig. 5, the effective exponent $x$ of the corrections to
scaling has been estimated to be close to one ($x \approx 0.95$), in
the range $0.02 <t< 0.1$.\\
Moving from the surface into
the bulk, the effective exponent is usually lowered, see Fig. 4. Eventually,
one may observe an interesting crossover phenomenon, as shown in
Fig. 5. Away from criticality, $T_c$, and
sufficiently deep in the bulk (e.g. $z= 10$), the effective
exponent follows closely the behaviour of the effective bulk critical
exponent, as obtained from the corresponding magnetization data 
of the Ising model with full periodic boundary conditions [12].
However, on approach to $T_c$, when the bulk correlation length becomes
comparable to the distance from the surface (= 10 lattice spacings, in
the example), $\beta_{eff}$ crosses over to its surface value. Indeed, at
each distance (large, but finite) from the surface, one reaches, 
sufficiently close to $T_c$, the surface exponent $\beta_1 \approx 0.80$
instead of the bulk exponent $\beta \approx 0.32$.\\ 
The susceptibility per layer, $\chi(z)$, displays close to $T_c$, at
the ordinary transition, $r=1.0$, a non--monotonic behaviour, see Fig.\ 6, with
the maximal value shifting towards the center of the system as the 
temperature gets larger. In analogy to (5), one may define an effective
exponent $\gamma_{eff} (z)$. In the bulk, that exponent is clearly positive,
with a value around one at $t \approx 0.1$, rising quite sharply
to at least about 1.5 on further approach to $T_c$
(finally, finite--size effects tend to decrease
the effective exponent again). However, $\chi$ remains, at $T_c$, finite at 
the surface, $z=1$, with a cusp-like singularity, i.e.
a negative critical exponent $\gamma_{11}$, by approaching criticality
from high temperatures [1,2]. To estimate $\gamma_{11}$, we
determined the temperature $T_m$ at which, for a finite system $(K, L, M)$,
$\chi_{11}$ exhibits its maximum, $\chi_{11}^m$. A suitably
defined effective exponent is

\be 
(\gamma_{11})_{eff}(t_m)= -d\ln (\chi_{11}^m - \chi_{11}(t_m))/d\ln (t_m)
\ee
 
\noindent
with $t_m$= $|T-T_m|/T_m$. Obviously, $(\gamma_{11})_{eff}
\longrightarrow \gamma_{11}$ as $K, L, M \longrightarrow \infty$,  
$T_m \longrightarrow T_c$, and $t_m \longrightarrow \ t   
\longrightarrow 0$. Based on our Monte Carlo data at $r=1.0$, this ansatz
leads to a rather rough estimate of $\gamma_{11}$, circumventing finite--size
effects again, with $\gamma_{11} = - 0.25 \pm 0.1$. This value may be
compared to results of previous simulations [4], where only the sign
of $\gamma_{11}$ was identified, and to predictions of renormalization
group calculations [1,2,6,7], $\gamma_{11} \approx -0.33$.\\
The critical exponent for the response function of the surface
magnetization to a bulk field, $\chi_1$, may be estimated in a more
standard and straightforward way, by computing in analogy to (5)
the effective exponent 

\be
(\gamma_1)_{eff}(t) = -d\ln (\chi_1(t))/d\ln (t) 
\ee
  
\noindent
At $r=1$, that exponent decreases with t, on approach to $T_c$, 
approximately linearly in the range $0.02 <t < 0.1$. A linear
extrapolation yields the asymptotic exponent $\gamma_1 = 0.78 \pm 0.05$,
in agreement with previous estimates [1,2,4].\\
We performed additional, albeit much less extensive simulations near
the special point, $ r=r_c$. In
particular, at $r=1.50$, $\beta_{eff} (z=1)$
is found to increase more strongly than linearly, at least down to $t =0.02$,
so that an extrapolation to $t \longrightarrow 0$ is not obvious. 
A reasonable estimate seems to be $\beta_1 =0.23 \pm 0.01$, in 
good agreement with a fairly recent large scale 
Monte Carlo study, $\beta_1 = 0.237 \pm 0.005$ [5], but a little bit
lower than the field theoretical value, $\beta_1 \approx 0.26$ [7].
Similarly, our estimate
for $\gamma_1 (= 1.5 \pm 0.1)$ is slightly higher than that suggested
by a renormalization group calculation, $\gamma_1 \approx 1.30$ [7]. In any
event, the agreement is reasonably satisfactory.\\  
Beyond the special point, $r > r_c$, the critical exponents at the
surface transition, $T_s > T_c$, are believed to be in the universality
class of the two--dimensional Ising model [1,2]. Indeed, our 
Monte Carlo data are consistent with that prediction.-- At this point, we
draw attention to recent Monte Carlo simulations on
short-range correlation functions near 
the surface transition, $T_s$, which have been performed to interpret 
spin-polarized photoelectron data of some magnets [15].

\section{Surface with random couplings}
To mimic an amorphous surface, possibly due to (non-)magnetic impurities
at the surface of the crystal, we replace the unique surface coupling
$J_s$ by two random, ferromagnetic interactions, being either
'strong', $J_{s1}$, or 'weak', $J_{s2}$, i.e. $J_{s1} > J_{s2} > 0$, see
Fig.\ 1b.-- Note that an alternate form of randomness, namely random 
surface fields, had been considered before
using Monte Carlo techniques [16].\\
Following detailed simulations of the two-dimensional dilute Ising
model [17,18], both couplings were assumed to occur with the same
probability. Then the ratio $d=J_{s2}/J_{s1}$ measures the degree of
dilution: $d =1$ corresponds to the perfect case, and $d =0$ to the
percolation limit. We performed most of the simulations at $d =1/10$,
where critical dilution effects are expected to be easily detectable, because
the crossover length to the dilution dominated critical regime is
only a few lattice constants [17]. The ratio of the surface to bulk 
couplings $r = (J_{s1} + J_{s2})/ (2 J_b)$
was varied in between 1.0 and 3.5, to study critical phenomena at the
ordinary transition and to locate the special point, $r =r_c(d)$.
For instance, $r =1.0$ is realized
when $J_{s1} = (20/11) J_b$ and $J_{s2} = (2/11) J_b$,
for $d= 1/10$. The sizes of Monte Carlo systems and lengths of
runs were chosen like in the perfect case.\\  
Because the critical temperature (thence, the effective 
interaction) of the
corresponding two--dimensional system is reduced by randomness, at
given mean coupling $(J_{s1} + J_{s2})/2$ [17], the special point may be
argued to be shifted towards larger ratios $r_c$, as dilution 
is increased (indeed, at $d =1/10$, for example, we
locate $r_c$ at $1.70 \pm 0.1$, see below, to be compared with 1.50 for
the perfect surface). Accordingly, at  $r =1.0$ and $d =1/10$ the ordinary
transition is encountered. We shall discuss our findings for that case
first, comparing it to results for the perfect surface, $d=1$.
Certainly, the transition temperature $T_c$ is not affected by the
randomness in the surface couplings.\\
At fixed temperature, $T < T_c$, the dilution tends to decrease the
magnetization $m(z)$ at and near the surface, for distances small
compared to the bulk correlation length. This trend is shown in Fig. 7
for the surface magnetization, $m_1$. At first sight, perhaps, somewhat
surprisingly, the energy per layer $E(z)$ is also lowered compared
to the perfect
situation, but only in the surface layer. This behaviour may be explained
by the ordering in clusters of strongly
interacting, $J_{s1} (> J_s)$, spins, with a weak coupling between
these, possibly oppositely oriented, clusters.\\
Despite the obvious drop in $m_1$ due to the surface randomness, the
ratio of $m_1(d=1/10)/ m_1(d=1)$ is almost constant, about 0.9, over
a wide range of temperatures, from $k_BT/J_b \approx  3.4$ up to $T_c$, with
a very shallow minimum near $k_BT/J_b = 3.8$. Therefore, the
effective critical exponent $\beta_{eff}(z=1)$, see Eq. (5),
follows very closely that of the perfect case, as shown in Fig. 8, leading 
to the same estimate for $\beta_1 = 0.80 \pm 0.01$, obtained from a linear
extrapolation towards $T_c$ of the data
for the effective exponent. Thence, the
randomness in the surface couplings seems to be irrelevant for
the asymptotic behaviour of the surface magnetization, and of minor
importance even for the corrections to scaling.
Similarly, the estimate for $\gamma_1$ is compatible with the 
one in the perfect case. Obviously, 
the numerical findings provide support to the conjecture
by Diehl and N\"usser, suggesting
that short-range correlated dilution of surface interactions is
irrelevant for surface critical exponents, based on a Harris--type
criterion [19] and perturbation-theory to first--order [10].-- The 
findings are corroborated by exact and simulational results
on two--dimensional Ising models with surfaces, where $\beta_1$ is
1/2, independent of surface dilution [20,21].-- Interestingly enough, in
three--dimensional Ising models, random
surface fields seem to be irrelevant as well [16].\\ 
To locate the special point, $r=r_c$, one may determine the surface
transition temperature, $T_s$. At
$ r < r_c$, $T_s = T_c \approx 4.5115 \, J_b/k_B$,
while $T_s(r) > T_c$ otherwise. Standard procedures may be applied, e.g.,
by analysing the finite--size dependence of the turning point in
$m_1$. The resulting phase diagram near the special point, for $d=1/10$,
is depicted in Fig.\ 9.\\
However, close to the special point, $T_s$ deviates only minutely
from $T_c$, and ambiguities may arise. To get easily a lower bound
for $r_c$, one may monitor effective exponents, e.g., $\beta_{eff}(z=1)$. For
instance, in the random case with $d=1/10$, at $r=1.53$, i.e. 
slightly above $r_c(d=1)$, $\beta_{eff}$ rises sharply
at $t < 0.05$, being consistent with the asymptotic value $\beta_1
\approx 0.80$. Even at $r=1.7$, the effective exponent still tends to increase
on approach to $T_c$, to a value close to that of the perfect surface
at the special transition, $\beta_1 \approx 0.23$.-- An 
estimate for an upper bound for $r_c$ may be obtained by assuming $T_s = T_c$,
and looking for inconsistencies in the critical exponents. In particular, such
an inconsistency occurs when $\beta_{eff}(z=1)$ becomes smaller than 1/8. 
Based on that kind of analysis, in addition to the standard procedures,
we reached at an upper bound $r_c < 1.8$.
In summary, a reasonable estimate of $r_c(d=1/10)$
seems to be $1.70 \pm 0.1$. To
refine it, more sophisticated methods are needed [5], which, however, are
beyond the scope of this study. The important point is, to notice that
$r_c$ shifts towards a higher value when diluting
the surface interactions.

\section{Surface with steps}
To study effects of corrugation on surface critical phenomena,
we shall consider here thermal properties at a step of monoatomic
height, superimposed on a perfect surface.\\
As in the perfect
case, we consider systems with $K \times L$ spins per $(x,y)$ layer, with
the exception of 
the topmost layer, $z=0$, where spins are restricted to the center
half, forming a strip-like terrace of monoatomic height along the 
$y$--direction, with two bordering steps, see Fig. 1c. At
the bottom, $z= M$, the surface is assumed to be flat.\\
Three kinds of interactions may be introduced. Spins at the step-edges are
coupled ferromagnetically, $J_e > 0$. Other spins at the top (or bottom)
of the system, having no upper (lower) neighbouring spins, interact
through the surface coupling $J_s > 0$, while all the remaining spins are
coupled by $J_b > 0$. The phase diagram in the ($r= J_s/J_b, k_BT/J_b$)--plane
is expected to be identical to that of the perfect surface [1], because the
couplings at the one--dimensional step--edges do not support additional
long--range ordering.\\
To quantify the influence of the additional terrace, we calculated
the magnetization per row

\be
m(x,z)= \frac{1}{L} \left< \left| \sum\limits_y  S_{xyz} \right| \right>.
\ee

\noindent
The magnetization at the step--edges, say, $x= x_{s1}, x_{s2}$,
$m_{11}= m(x_{s1},z=0)
= m(x_{s2},z=0)$ is of particular
interest, deviating possibly most significantly 
from the magnetization of the perfect surface, $m_1$. Certainly, the
aim is to identify the behaviour at a single step, i.e.
the width of the terrace
has to be large compared to the relevant correlation lengths, as it is
the case in the limit of the semi--infinite system, 
($K, L, M \longrightarrow \infty$ ). Therefore, careful finite--size
analyses are needed again, considering the effect of all three linear
dimensions, $K, L, M$, on the quantities of interest, especially $m_{11}$.
For instance, on approach to criticality, $T_c$, the length
of the step, $L$, is of importance [21]. To monitor and, eventually,
circumvent finite--size effects, $K$ ranged from 80 to 120, $L$ from
80 to 240, and $M$ from 40 to 80. As before, we used the one--cluster--flip
Monte Carlo algorithm, generating a few $10^5$ clusters, and averaging
typically over about 5 realizations.\\
In Fig. 10, the magnetization per row, $m(x,z)$, is depicted at fixed
temperature near the ordinary transition, taking $J_e= J_s = J_b$, for
various layers, starting at the top, $z=0$, and moving into the bulk.
The magnetization is minimal at the step--edge, due to
the reduced coordination number at the step.
To avoid finite--size effects, the system size has to be chosen
in such a way, that, e.g., $m(x,z)$ acquires (i) the numerically
accurately known bulk magnetization [11,12] sufficiently
far away from the surface, and (ii) the correct
surface magnetization, $m_1$, at the bottom,
$m(x,M)$, as determined in the perfect case. Furthermore,
the influence of the length of the step, $L$, on the step--edge
magnetization needs to be
scrutinized, thereby monitoring the impact of bulk and surface correlation
lengths.\\
The effective exponent of the step--edge magnetization,
$(\beta_{eff})_{11}(t)$, defined in complete analogy to (5), is shown
in Fig. 11. In the figure, Monte Carlo data affected by finite--size
effects are included, characterised by a decrease in the
effective exponent as one goes closer to $T_c$. As observed readily, 
finite--size effects play an important role for unusually large
systems even quite far away from $T_c$.\\
In the following we
consider only the data unperturbed by finite--size effects.
Comparing to Fig. 5, one notices an enhancement of the exponent
$(\beta_{eff})_{11}(t)$,
at fixed distance from $T_c$, i.e. at fixed $t$, relative
to the effective exponent
of the surface magnetization. In any event, $(\beta_{eff})_{11}(t)$
seems to vary on approach to criticality almost
linearly. Based on a linear extrapolation,
the asymptotic value of $\beta_{11}$ may be estimated to be $0.80 \pm 0.015$,
in agreement with that of the magnetization of a flat surface.
In contrast to the case of random couplings, corrections to scaling
of the magnetization at the step--edge
are distinctively different from those of $m_1$ for the perfect surface.\\
Our result for $\beta_{11}$ on surfaces with steps may be argued to
be in accordance with the conjecture of Diehl and N\"usser, stating
that smooth corrugations correspond to irrelevant perturbations, in the
sense of the renormalization group theory [10]. The conjecture, however,
refers merely to the critical behaviour of the entire surface, and may
not necessarily apply to the step--edge magnetization.

\section{Summary}
Using Monte Carlo techniques, we studied critical properties at the
ordinary transition, the special point, and the distinct surface 
transition of semi--infinite Ising systems on cubic lattices with
nearest--neighbour
ferromagnetic interactions. In particular, the influence of randomness
and corrugations on the critical exponents at the ordinary transition
has been investigated.\\
Especially the critical exponent of the surface magnetization has been
shown to be robust against both types of perturbations, as mimicked
by random 'strong' and 'weak' couplings in the surface, and by
steps of monoatomic height. Its value is always
$0.80 \pm 0.01(5)$. Even corrections to scaling are affected only rather
mildly by the dilution, in contrast to the corrugated case comparing
the step--edge magnetization to the surface magnetization, where
the correction terms to the asymptotic power--law are found to differ
appreciably.\\
Our Monte Carlo findings support and extend conjectures of Diehl
and N\"usser, based on renormalization group arguments, suggesting
the irrelevant character of these surface perturbations at the
ordinary transition. The findings may also be helpful in
interpreting experiments on surface critical phenomena, by
showing that different 'dirt effects' are of minor importance
for the asymptotic critical exponents.\\[1.0cm]

\noindent{\large \bf Acknowledgements}\\[2ex]
 
\noindent
We should like to thank K. Binder, H. W. Diehl, S. Dietrich, E. Eisenriegler,
D. P. Landau, and U. Ritschel for very useful discussions. The interactions
with them at the 'Landau-Seminar' in Bad Honnef and at the CECAM-Workshop
in Lyon, September 1997, have been most helpful.

\vskip 1.3cm
\bc
{\Large \bf References}\\[2ex]
\ec
\begin{enumerate}
\item Binder, K.: In: Phase Transitions and Critical Phenomena. Vol.8,
Domb, C. and Lebowitz, J.L. (eds). London: Academic Press 1983
\item Diehl, H.W.: In: Phase Transitions and Critical Phenomena. Vol.10,
Domb, C. and Lebowitz, J.L. (eds). London: Academic Press 1986; In:
Proceedings of Third International Conference 'Renormalization Group-96'.
Singapore: World Scientific (to be published)
\item Dosch, H: Critical Phenomena at Surfaces and Interfaces. Berlin,
Heidelberg, New York: Springer 1992
\item Landau, D.P., Binder, K.: Phys. Rev. B {\bf 41}, 4633 (1990)
\item Ruge, C., Dunkelmann, S., Wagner, F., Wulf, J.: J. Stat. Phys.
{\bf 73}, 293 (1993); Ruge, C., Dunkelmann, S., Wagner, F.:
Phys. Rev. Lett. {\bf 69}, 2465 (1992)
\item Diehl, H.W., Dietrich, S.: Z. Physik B {\bf 42}, 65 (1981)
\item Diehl, H. W., Shpot, M.: Phys. Rev. Lett.{\bf 73}, 3431 (1994)
\item Krimmel, S., Donner, W., Nickel, B., Dosch, H., Sutter, C.,
Gr\"ubel, G.: Phys. Rev. Lett. {\bf 78}, 3880 (1997); see also
Ritschel, U.: Preprint (1997)
\item Kaneyoshi, T.: Introduction to Surface Magnetism. Boca Raton, Ann
Arbor, Boston: CRC Press 1991
\item Diehl, H.W., N\"usser, A.: Z. Physik B {\bf 79}, 69 (1990)
\item Ferrenberg, A.M., Landau, D.P. :Phys. Rev. B {\bf 44}, 5081 (1991)
\item Talapov, A. L., Bl\"ote, H. W. J.: J. Phys. A: Math. Gen. {\bf 29}, 5727
(1996)
\item Binder, K., Landau, D.P.: Phys. Rev. B {\bf 37}, 1745 (1988)
\item Wang, J. S., Swendsen, R.H.: Physica A {\bf 167}, 565 (1990);
Wolff, U.: Phys. Rev. Lett. {\bf 60}, 1461 (1988)
\item Zhang, F., Thevuthasan, S., Scalettar, R.T., Singh, R.R.P., 
Fadley, C.S.: Phys. Rev. B {\bf 51}, 12468 (1995)
\item Mon, K.K., Nightingale, M.P.: Phys. Rev. B {\bf 37}, 3815 (1988)
\item Wang, J. S., Selke, W., Dotsenko, Vl. S., Andreichenko, V.B.:
Physica A {\bf 164}, 221 (1990)
\item Selke, W., Shchur, L. N., Talapov, A. L.: In: Annual Reviews of
Computational Physics. Vol. 1, Stauffer, D. (ed). Singapore: World
Scientific 1994
\item Harris, A. B.: J. Phys. C {\bf 7}, 1671 (1974)
\item Mc Coy, B. M., Wu, T. T.: The Two--Dimensional Ising Model. Cambridge:
Harvard University Press (1973)
\item Selke, W., Szalma, F., Lajko, P., Igloi, F.: J. Stat. Phys., in 
print (1997)
\end{enumerate}

\bc
{\Large \bf Figure Captions}\\[2ex]
\ec
\begin{itemize}
\item[Fig. 1:] Geometry and interactions of the Ising models 
for (a) the perfect surface,
and surfaces with (b) random couplings and (c) steps.
 
\item[Fig. 2:] Schematic phase diagram for the semi--infinite
three--dimensional Ising
model; $r$ denotes the ratio of the surface to the bulk couplings. 
 
\item[Fig. 3:] Magnetization profiles $m(z)$ for the perfect
surface with $J_s = J_b$ at, from bottom to top, $k_BT/J_b$ =
4.49, 4.45, and 4.35, for systems of size $100 \times 100 \times 80$. The
dashed line denotes the bulk value [12].
 
\item[Fig. 4:] Effective exponent of the magnetization per
layer, $\beta_{eff}(z,t)$, of the Ising model with a
perfect surface, with $r= 1.0$ at, from
bottom to top, $k_BT/J_b$ =4.25, 4.425, and 4.46, for systems of size
$50 \times 50 \times 40$.
 
\item[Fig. 5:] $\beta_{eff}(z,t)$ in the perfect case, with $r= 1.0$, at
fixed distances from the surface, $z$, as a function of reduced
temperature $t$, for systems of size $25 \times 25 \times 20$ 
(triangles), $50 \times 50 \times 40$ (diamonds), 
$100 \times 100 \times 80$ (squares), and
$150 \times 150 \times 120$ (circles). The dashed line follows from
bulk Monte Carlo data [12].
 
\item[Fig. 6:] Susceptibility per layer $\chi(z)$ in the perfect case
with $r= 1.0$ at, from bottom to
top, $k_BT/J_b$ = 4.2, 4.35, 4.45, and 4.47,
for systems of size $50 \times 50 \times 40$.
 
\item[Fig. 7:] Surface magnetization $m_1$ vs. temperature, for
the perfect and random surface, at $r =1.0$, for systems of size 
$50 \times 50 \times 40$.
 
\item[Fig. 8:] Effective exponent of the surface
magnetization, $\beta_{eff}(z=1)$, for a surface with random couplings,
with $J_{s2} = J_{s1}/10$ and $r= 1.0$. 
 
\item[Fig. 9:] Phase diagram for the three-dimensional Ising model with
random couplings in the surface, $J_{s2} = J_{s1}/10$, close
to the special point (at $r_c = 1.70 \pm 0.1$, as indicated in the
figure).
 
\item[Fig. 10:] Profiles of the magnetization per row $m(x,z)$
for the Ising model with a terrace of monoatomic height on the surface,
with $J_e = J_s = J_b$, at $T= 0.95 T_c$, for systems of size
$80 \times 80 \times 40$. 
 
\item[Fig. 11:] Effective exponent of the step--edge
magnetization, $(\beta_{eff})_{11}(t)$, as a function of reduced
temperature $t$, with $J_e = J_s = J_b$, for various system
sizes.

\end{itemize}

\end{document}